\documentclass[%
reprint,
superscriptaddress,
 amsmath,amssymb,
aip,
]{revtex4-1}

\usepackage{comment}
\usepackage[dvipdfmx]{graphicx}
\usepackage{dcolumn}
\usepackage{bm}

\begin{document}


\title{Dynamic Evolution of Flux Distributions in a Pulse-driven Superconductor by High-speed Magneto-optical Imaging }

\author{H. Kurokawa}
\affiliation{ Department of Basic Science, The University of Tokyo, 3-8-1, Komaba, Meguro-ku, Tokyo 
153-8902, Japan}
 \author{Y. Kinoshita}
 \affiliation{The Institute for Solid State Physics, The University of Tokyo, 5-1-5, kashiwanoha, 
 	kashiwa-shi, Chiba 277-8581, Japan}
 \author{F. Nabeshima}
\affiliation{ Department of Basic Science, The University of Tokyo, 3-8-1, Komaba, Meguro-ku, Tokyo 
153-8902, Japan}
 \author{M. Tokunaga}
 \affiliation{The Institute for Solid State Physics, The University of Tokyo, 5-1-5, kashiwanoha, 
 	kashiwa-shi, Chiba 277-8581, Japan}
 \author{A. Maeda}
\affiliation{ Department of Basic Science, The University of Tokyo, 3-8-1, Komaba, Meguro-ku, Tokyo 
153-8902, Japan}

\date{\today}

\begin{abstract}
	
	The accurate understanding of flux dynamics is essential for the design and operation of superconducting circuits. The time evolution of flux-density distribution in an NbN strip by the transport current was observed using high-speed magneto-optical microscopy. It was determined that even for the dynamic penetration and exclusion of vortices under the transport current, the surface barrier is essential.  This feature is important for the correct understanding of the complex behavior of state-of-the-art superconducting devices.

		
\end{abstract}

\maketitle


Recently, superconducting circuits have become increasingly important for the 
quantum computing,\cite{Nakamura1999,Clarke2008a,Devoret2013}, ultra-high sensitivity sensors,\cite{Irwin1996,Vayonakis2003,Bal2012,Lachance-Quirion2019} 
and circuit quantum electrodynamics.\cite{AndreasWallraff2004,Xiang2012} The transport current and magnetic field are often introduced to control the circuits,\cite{Healey2008,Chen2011,Bothner2012,Kroll2019} and the circuit responses are usually determined by the kinetic inductance of the superfluid \cite{Gittleman1965,Vissers2015} and the motion of vortices.\cite{Bothner2012,Kwon2018a,Kurokawa2019} However, the current and flux are  highly non-uniformly distributed in a thin strip, which is one of the most common forms  of the circuit line, owing to the pinning, sample geometry, and surface barrier.\cite{Brandt1993,Zeldov1994a,Zeldov1994,Ginodman1999} This non-uniformity sometimes leads to complex and hysteretic behaviors of the devices. Thus, the accurate understanding of the dynamic current and flux distributions is essential to correctly understand the behaviors of these devices. 

The critical-state model (CSM) is the fundamental model describing the current and flux-density distribution in a superconductor under magnetic field in the presence of pinning.\cite{Bean1962,Swartz1968,Brandt1993,Zeldov1994a} In addition, it has been revealed that the surface of the superconductor acts as the barrier for the flux entry, which is effective even in the absence of bulk pinning (surface barrier).  The existence of surface barrier considerably changes the current and flux distribution,\cite{Bean1954,Zeldov1994,Plourde2001} which results in the delay in flux penetration, hysteretic magnetization,\cite{Zeldov1994} and enhancement of critical current, $I_\textrm{c}$.\cite{Jones2010} 
Several mechanisms have been suggested as the origin of the surface barrier.\cite{Bean1954,Zeldov1994,Plourde2001} One of the most representative mechanisms is the Bean--Livingston (BL) barrier, which is the result of the competition between two forces acting on the vortex near a parallel surface. The first force is an inward force owing to the presence of shielding current: the other force is an outward force owing to the attraction between the vortex and its mirror image outside the sample.\cite{Bean1954} Another example is the geometrical barrier, which is the product  of sample geometry.\cite{Zeldov1994} Because the vortex bends at the sample edge of the flat rectangular sample, the potential energy of the vortex varies across the sample, which acts as the barrier for the entry of the vortex. 
The ability to directly image magnetic field and current distribution is a powerful tool that will allow to investigate the effect of surface barrier because it can explicitly separate the contribution of the bulk from that of the edge. \cite{Fuchs1998, Grigorenko2001a, Gutierrez2013}
A study using a micro Hall-sensor array has revealed that most parts of the current flow at the edges of the sample,\cite{Fuchs1998} and the enhancement of the critical current at the edges was observed using the space-resolving low-temperature laser scanning microscopy method.\cite{Sivakov2018} 

The vortex system often shows metastable transient states owing to the bulk pinning, thermal fluctuation, and surface barrier,\cite{Giller2000a,Bobyl2002, Wells2016} which depend on the current or magnetic field history. Thus, in addition to spatial imaging, the imaging of vortex dynamics in the time domain in the presence of driving current is  necessary for the accurate understanding of complex hysteretic and metastable behaviors of superconducting devices.\cite{Lahl2003,Bothner2012,Kwon2018,Kwon2018a, Kurokawa2019} 

 In this letter, we show the flux dynamics in an NbN film in the spatially and time-resolved manner using the high-speed magneto-optical microscopy in the presence of transport current. 
 The processes of flux penetration and exclusion were monitored for three qualitatively different initial flux distributions. Using this technique, we succeeded in revealing the effects of the surface barrier on the dynamic flux entry and exit in the time-domain. It is determined that even for the dynamic behavior of the flux-density distribution under the transport current, the surface barrier is essential.



\begin{figure}
	\centering
	\includegraphics[width=1\linewidth]{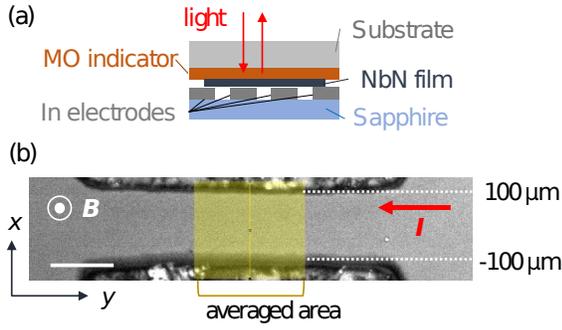}
	\caption{(a) Schematic diagram of the cross section of the magneto-optical imaging apparatus. (b) Optical image of the NbN strip. The light intensities in the yellow square were averaged along the $y$ direction. The red  arrow shows the current flow direction. The white scale bar is 200 $\mu$m. }
	\label{fig:figure1}
\end{figure}

Figure \ref{fig:figure1} (a) shows the schematic diagram of magneto-optical imaging. An NbN film was deposited on a magneto-optical (MO) indicator (Bi-substituted garnet film) by 
radio-frequency sputtering. The garnet film had an in-plane magnetization. The thickness of the NbN film was 1 $\mu$m, and the 
transition temperature was 15.1 K. The strip (Fig.\ref{fig:figure1} (b)) was formed by placing a metal mask on the garnet film before sputtering. The NbN film on the MO indicator was attached to a 1-mm-thick sapphire substrate using indium, which was also used as electrodes. 
The sapphire substrate was mounted onto a cold finger of a refrigerator using a silver paste. To monitor the temperature of the sample, a thermometer (CX-SD, Lake Shore) was attached to the top of the sample. The temperature of the sample was kept at 11 K. Magnetic field was applied perpendicular to the film using a copper coil.  


To drive vortices, a triangular current pulse was applied to the NbN film by the combination of a function generator (3390, Keithley) and an $I$--$V$ converter (BWS18-15, Takasago). The width of the one-shot pulse was 5 ms. The applied current was monitored by measuring the voltage drop of a 100 m$\Omega$ resistor in series with the sample using an oscilloscope (DL750, YOKOGAWA).  
The voltage induced by the flux motion in the sample was simultaneously measured and monitored by the oscilloscope. 

The changes in the flux density by the applied current were observed as follows. The linearly polarized incident light was reflected at the surface of the NbN film. Magnetic flux parallel to the incident light induces a rotation in the polarization vector of light owing to the Faraday effect. The reflected light was focused by the objective lens (5$\times$) and passed through the second polarizer.  The angle between the first and second polarizers was set at 10$^\circ$ from the crossed Nicole configuration. The reflected light was detected by a high-speed camera (Q1V, NAC Image Technology), and the frame rate was 10000 frames per second. The other details about optics and cryogenics are described in Ref. \onlinecite{Katakura2010}.
The absolute value of the magnetic field at the edges could not be calculated owing to the non-uniform reflection of the light at the edges, which made the calibration impossible at the edges in the sample. However, the changes in the flux density could be estimated from the changes in the light intensity, $i$, because $\Delta i(t)/i_0  \equiv (i(t)-i(t=0))/i(t=0)$ is proportional to the change in the flux density, $B(t)-B(t=0)$, where $t$ is the elapsed time after triggering the application of current. To investigate the effect of current on different magnetic field profiles, we prepared for the field-cooled (FC) state, zero-field-cooled (ZFC) state, and remanent state, where the magnetic field was decreased to zero after field-cooling (10.5 mT). 
In the case of FC and ZFC, an external magnetic field of 10.5 mT was applied during the imaging experiment.

\begin{figure}[t]
	\centering
	\includegraphics[width=1\linewidth]{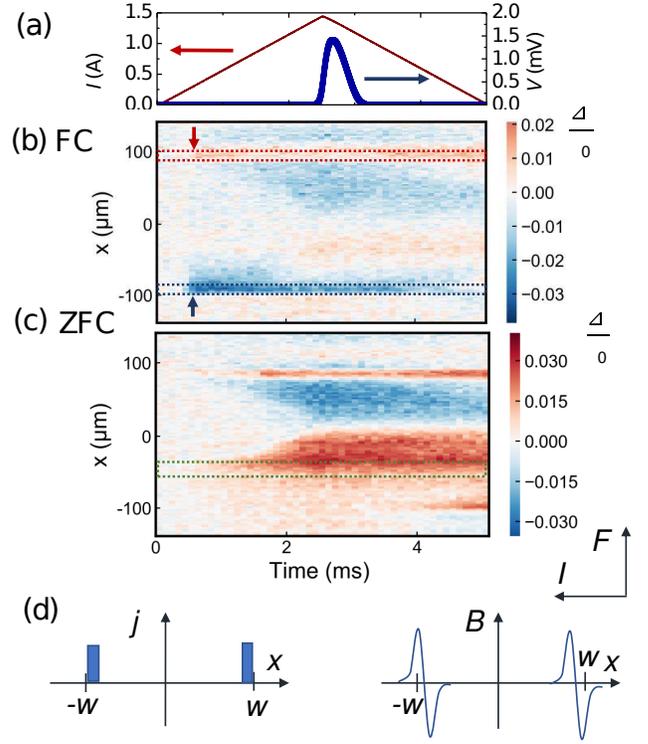}
	\caption{(a) Applied current as a function of time (red line), and the voltage induced by the flow of vortices (blue line). (b) Normalized local intensity change as a function of time in the field-cooled (FC) state. The red and blue dotted frames enclose the areas where sudden intensity changes were observed. (c) Normalized local intensity change in the zero-field-cooled (ZFC) state. The green dotted frame encloses the area where the intensity started to increase at the initial stage inside the strip. The directions of the applied current and driving force are shown at the bottom right of the figure. (d) Schematics of the current and flux-density distribution assuming that the applied current flowed at the edges.
	}
	\label{fig:figure2}
\end{figure}

Figure \ref{fig:figure2} (a) shows the applied current as a function of time with the voltage drop induced by the vortex motion in the FC state.  The critical current, $I_\textrm{c}$, was 
estimated as 1.27 A with the criteria of 10 $\mu$V, and the corresponding critical current 
density, $j_\textrm{c}$, was 6.35$\times10^5$ A/cm$^2$. 

\begin{figure}[t]
	\centering
	\includegraphics[width=1\linewidth]{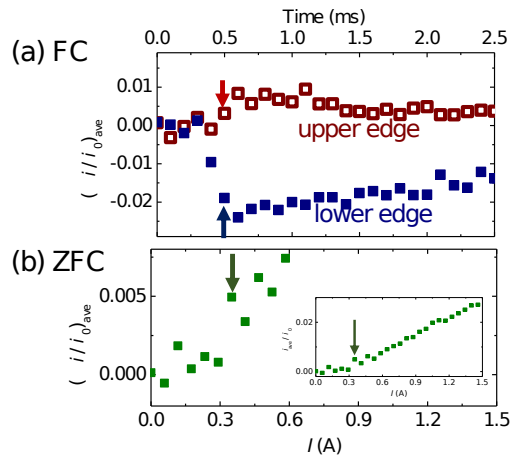}
	\caption{(a) Average $\Delta i(t, x)/i_0(x)$ of the FC state as a function of current or time. $\Delta i(t, x)/i_0(x)$ was averaged over 88 $\mu$m $<x<$ 100 $\mu$m for the upper edge (red frame in Fig. \ref{fig:figure2} (b)) and $-$100 $\mu$m $<x<$ $-$88 $\mu$m for the lower edge (blue frame in Fig. \ref{fig:figure2} (b)).	The red and blue arrows correspond to 0.3 A. 
	(b)  Average $\Delta i(t, x)/i_0(x)$ of the ZFC state as a function of the current or time. The average was obtained for $-$50 $\mu$m $<x<$ $-$28 $\mu$m (green frame in Fig. \ref{fig:figure2} (c)). The inset shows the overall view. The green arrows point to the square at which the average $\Delta i(t, x)/i_0(x)$ increased. }
	\label{fig:AveInt}
\end{figure}

Figure \ref{fig:figure2} (b) shows $\Delta i(t, x)/i_0(x)$ of the FC state in the time vs. $x$ plane, where $x$ is the position. The direction of the driving force was from the bottom to the top. The flux density increased in the red region, while it decreased in the blue region. 
The flux density suddenly increased (decreased) at the upper (lower) edge of the film at approximately 0.3 A (red and blue arrows in Fig. \ref{fig:figure2} (b)). For clarity, the intensity at the upper edge (red frame in Fig. \ref{fig:figure2} (b), 88 $\mu$m $<x<$ 100 $\mu$m ) and lower edge (blue frame in Fig. \ref{fig:figure2} (b), -100 $\mu$m $<x<$ -88 $\mu$m ) was averaged and plotted as a function of time and current in Fig. \ref{fig:AveInt} (a). It was clearly observed that the average intensity increased at approximately 0.3 A at the upper edge while it decreased at the lower edge. At larger currents, the average intensity tended to saturate at both edges. 

One of the most remarkable features is the presence of the threshold current value for the vortex entry ($\sim$ 0.3 A in Fig. \ref{fig:figure2} (b)), which strongly suggests that the surface barrier is essential.
In the presence of the surface barrier and the pinning, most of the current flows at the edges of the strip.\cite{Zeldov1994, Beidenkopf2009} 
In this picture, as soon as the self-field at the edges became larger than the first flux penetration field, $H_\textrm{p}$, the flux owing to the self-field started to enter the edges. Thus, the flux penetration at approximately 0.3 A was considered to correspond to this process. 
In addition, another indication of the presence of the surface barrier is the opposite sign of the change in the flux density at the flux entry between the upper edge (red) and lower edge (blue). Assuming that the current flowed primarily at the edges, the direction of the self-field inside the strip is opposite in each edge of the strip, as shown in Fig. \ref{fig:figure2} (d).
  
We estimated the strength of the surface barrier from the data in Fig. \ref{fig:figure2} (b). 
First, suppose that only the bulk pinning and geometrical barrier exist.  In this case, the flux cannot penetrate the sample until the current density at the edge reaches the threshold current, $j_\textrm{E} = j_\textrm{E}^0 + j_\textrm{c}$, where $j_\textrm{E}^0= 2H'/d$, $d$ is the thickness of the film, and $H'$ is the characteristic field at which the first flux penetration occurs, and $j_\textrm{c}$ is the bulk critical current density.\cite{Zeldov1994} $H'$ is equal to the lower critical field, $H_\textrm{c1}$, in the absence of the BL barrier.\cite{Zeldov1994} In the presence of the BL barrier, $H'$ is replaced by the BL surface barrier penetration field,\cite{Zeldov1994} which is on order of the thermodynamic critical field, $H_\textrm{c}$.\cite{Burlachkov1993} 
For the first approximation, we assumed that the applied current density, $j_\textrm{a}(x)$, was zero in the center of the strip ($-$88 $\mu$m $<x<$ 88 $\mu$m), and $j_\textrm{a}(x)=j_\textrm{E}$ in the edges ($-$100 $\mu$m $<x<$ $-$88 $\mu$m and 88 $\mu$m $<x<$ 100 $\mu$m) where $\Delta i(t, x)/i_0(x)$ increased at 0.2--0.3 A. Then, $j_\textrm{E}$ and $j_\textrm{E}^0$ are determined to be 8.3--12.5 $\times10^5$ A/cm$^2$ and 2.0--6.2 $\times10^5$ A/cm$^2$, respectively. Then, $\mu_0H'$ becomes 1.3--3.9 mT, which was slightly smaller than the previously reported lower critical field, $\mu_0H_\textrm{c1}=$4--20 mT.\cite{Mathur1972, Beebe2016} 
Thus, this rough estimate shows that the geometrical barrier may play a role in preventing the flux penetration in the data of Fig. \ref{fig:figure2} (b). In addition, small $H'$ compared to the $\mu_0H_\textrm{c}\sim150 $ mT\cite{Mathur1972} suggests that the BL barrier was very weak even if it was effective.

    

After the sudden changes in the flux density at the edges ($\sim$0.5 ms, 0.3 A), the flux density started to slightly decrease in the upper side ( 0 $\mu$m $<x<$ 85 $\mu$m) as shown in Fig. \ref{fig:figure2} (b) (0.3--1.4 A), and the flux density decreased area (blue region) gradually increased and approached the center of the film. On the contrary, the flux density increased in the lower side ( $-$60 $\mu$m $<x<$ 0 $\mu$m), where the flux density increased area (red region) also approached the center of the film when the current increased. At larger currents ($>$1.27 A), the flux density changed area merged at the center of the strip. Then, the local current density, $j(x)$, became $j_\textrm{c}$ everywhere in the strip.\cite{Zeldov1994a} This corresponds to the appearance of the voltage induced by the macroscopic flux motion in Fig. \ref{fig:figure2} (a). Of note, the changes in the flux density were asymmetric to the center of the film, $x=0$, below 1.27 A. 
This asymmetry occurs owing to the penetration of the self-field induced by the transport current.   


Figure \ref{fig:figure2} (c) shows $\Delta i(t, x)/i_0(x)$ in the ZFC state. The flux density increased at the center of the film (red region in $-$50 $\mu$m $<x<$ 10 $\mu$m) when the applied current increased. An increase in the flux density at the center of the strip started at approximately 0.3 A. For clarity, we averaged $\Delta i(t, x)/i_0(x)$ in $-$50 $\mu$m $<x<$ $-$28 $\mu$m and plotted it in Fig. \ref{fig:AveInt} (b) similar to the FC case. A small increase in the averaged intensity was clearly observed at 0.35 A. In the ZFC state, the center of the sample was flux-free at low magnetic fields (applied after cooling).\cite{Zeldov1994a} Thus, the vortices entered the flux-free area with an increase in the applied current, and the flux density at the center ($-$50 $\mu$m $<x<$ 10 $\mu$m) increased.

The observed behaviors differ from what is expected in simple CSM, in which $j$ is assumed to be equal to $j_\textrm{c}$ everywhere.\cite{Zeldov1994a} Thus, as soon as the current is applied, $j$ surpasses $j_\textrm{c}$, and the flux smoothly penetrates inside the field-free area. However, the data in Fig. \ref{fig:AveInt} (b) shows that the flux penetration at the center of the strip started at 0.35 A. Of note, this number is almost the same as the flux entry field for the FC case (0.3 A). Thus, we consider that the delay of the flux penetration to the center in the ZFC case is also caused by the surface barrier. 
\begin{figure}[t]
	\centering
	\includegraphics[width=1\linewidth]{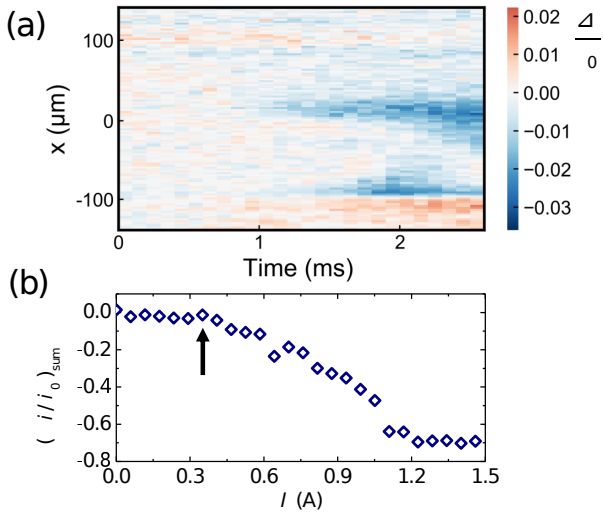}
	\caption{(a) Normalized local intensity change as a function of time in the remnant state. (b) Sum of $\Delta i(I, x)/i_0(x)$ inside the film ($-$100 $\mu$m $<$ $x$ $< $ 100 $\mu$m) as a function of applied current. The blue arrow corresponds to 0.35 A.}
	\label{fig:figure3}
\end{figure}

Thus far, we argued that the surface barrier was essential for the dynamic vortex entry.  To obtain further insight, we performed a  similar experiment for the remanent state in which the flux trapped inside the film forms the critical state. Thus, we observed the flux exclusion process by the application of current.  Figure \ref{fig:figure3} (a) shows the time dependence of $\Delta i(t, x)/i_0(x)$ in the remanent state.  The flux density started to decrease  with an increase in the current.  However, in Fig. \ref{fig:figure3} (a), it is not clear whether a finite threshold value exists for the flux exclusion.  Thus, we will discuss the sum of $\Delta i(t, x)/i_0(x)$, which corresponds to the change of the magnetization per unit volume. 

Magnetization per unit volume, $M\ (x)$, was equal to $B\ (x)/\mu_0$ in the film because the applied magnetic field was zero in the remanent state, where $\mu_0$ is the vacuum permeability. The normal component of $B$ is continuous at the boundary of the film and MO indicator. Thus, the $M\ (x)$ of the film was proportional to $B$ at the MO indicator, which was actually measured. Therefore, the change of the magnetization, $\Delta M (I)\equiv M\ (I)-M\ (I=0)$, is proportional to the sum of $\Delta i(I, x)/i_0(x)$ as
\begin{equation}
\begin{split}
\Delta M (I)  
&=\int B (I,x)-B (I=0,x)\textrm{d}x\\
&\propto   \int (i (I,x)-i (I=0, x)/i (I=0, x)\textrm{d}x.     
\end{split}
\end{equation}


Figure \ref{fig:figure3} (b) shows the sum of $\Delta i(I, x)/i_0(x)$ inside the film ($-$100 $\mu$m $<$ $x$ $< $ 100 $\mu$m). Although the simple CSM without the surface barrier expect the linear current dependence in $\Delta M\propto\int\Delta i(I, x)/i_0(x) \textrm{d}x$,\cite{Zeldov1994a} it was almost unchanged until 0.35 A, and then started to decrease. Thus, the existence of the finite threshold current became clear. The origin of the threshold is considered to be the surface barrier also in this case. The current flowed at the edges of the strip until the applied current reached a threshold, 0.35 A, above which it flowed in the center of the film, and the flux exclusion was assumed to start.

In summary, we observed the time evolution of the flux-density distribution in an NbN strip for the FC, ZFC, and remanent states under the presence of the driving current. We determined that the surface barrier was essential for the dynamic flux entry and exit in all cases. In addition, we successfully estimated in the time evolution of the current flowing area from the changes in the reflected light intensity. The current started to flow inside the strip only after the current density at the edge reached a threshold value. The obtained data is useful for the understanding the complex hysteretic and metastable behavior of superconducting circuits and improving the performance of state-of-the-art superconducting devices.

This work was supported by JSPS KAKENHI Grant Number JP16H00795. The devices were fabricated in the clean room for analog-digital superconductivity (CRAVITY) in the National Institute of Advanced Industrial Science and Technology (AIST).

\section*{DATA AVAILABILITY}
The data that supports the findings of this study are available within the article.

\section*{REFERENCES}

\end{document}